\documentstyle[twoside,fleqn,espcrc2]{article}

\input epsf
\def\GeV{\mathop{\rm GeV}\nolimits}
\def\MeV{\mathop{\rm MeV}\nolimits}
\title{On the $N_f$ and $a$ dependence of $B_K$}

\author{G. Kilcup, D. Pekurovsky and L. Venkataraman
        \address{Department of Physics, The Ohio State University, 
            174 W. 18th Ave., Columbus, Ohio 43210} } 
            
\begin{document}

\begin{abstract}
We present results of a study of $B_K$ using tadpole
improved gauge-invariant staggered operators. Using three
ensembles of $16^3 \times 32$ configurations with varying
numbers of dynamical flavors, we observe a small dependence
on $N_f$.  Using 7 quenched ensembles at different values
of $\beta$, we extrapolate to $a=0$.
\end{abstract}

\maketitle

\section{Introduction}
The $B_K$ parameter serves to parameterize the weak hadronic
matrix element responsible for $K^0-\bar{K^0}$ mixing.
Since this mixing gives us the only  CP violation
observed to date, $B_K$ is a crucial link between the measured 
quantity $\epsilon$ and the parameters of the Standard Model.
Lattice calculations are well suited for the study of $B_K$
parameter, and it has by now received much attention.
After an early round of calculations\cite{GKPS1,GKPS2,GKPS3},
the statistics have now been raised to a level which
allows one to examine some of the fine points of the calculation,
such as checks on the reliability of one-loop lattice perturbation
theory~\cite{Ishizuka93},
the chiral behavior and nondegenerate quark masses~\cite{Lee,Aoki96},
the dependence of $B_K$ on the lattice spacing~\cite{GKPS3,Aoki95,Aoki96}
and the number of dynamical flavors\cite{Kilcup93}.
In this note we offer more information on these latter two points.

\section{Calculational Setup}

\begin{table}[hbt]
\setlength{\tabcolsep}{1.2pc}
\caption{Ensembles for $N_f$ Study}
\label{tab:dynamical}
\begin{tabular}{llll}
\hline
$N_f$ & $\beta$ & $N_{\rm config}$ & $m_\rho a$ \\
\hline
0 & 6.05 & 306 & 0.384(5) \\
2 & 5.7 & 83 & 0.384(4) \\
4 & 5.4 & 69 & 0.391(7) \\
\hline
\end{tabular}
\end{table}

\begin{figure}[htb]
\begin{center}
\leavevmode
\epsfxsize 7.2cm 
\epsfbox{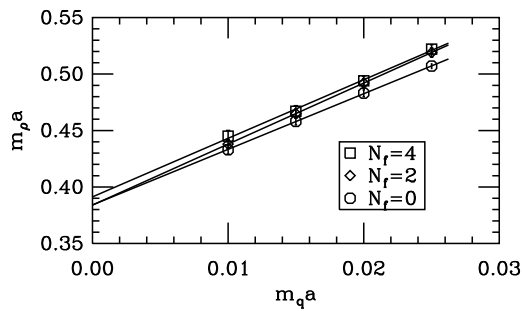}
\end{center}
\vskip-.2cm
\caption{Data and linear fit for $m_{\rho}a$ vs. quark mass, for 
three sets of configurations with $N_f$=0,2 and 4.
(See the talk by D. Chen [9].)
\label{fig:rho} }
\end{figure}

For the dynamical fermion comparison we use lattices of
geometry $16^3\times 32$, with parameters as given in
table \ref{tab:dynamical}.
The quenched configurations were generated on
the Ohio Supercomputer Center T3D, while the
dynamical configurations with two and four flavors of
$m_qa = 0.01$ staggered fermions were generated on the 256-node
Columbia machine.  
The parameters were chosen so as to make the scales of the
lattices exceedingly close (and equal to approximately
(2 GeV$)^{-1}$), as determined from the $\rho$-meson mass in
chiral limit (see Fig.~\ref{fig:rho} and Ref.~\cite{Chen}).
We employ 9 values of (degenerate) valence $d$ and $s$ quark masses
from $m_q=.01$ to $m_q=.05$.

For the continuum limit study we generated
7 ensembles of quenched configurations as listed
in table \ref{tab:quenched}, and used 7 to 9 values of $m_q$.

\begin{table}[htb]
\setlength{\tabcolsep}{0.7pc}
\caption{Quenched Ensembles for Continuum Extrapolation}
\label{tab:quenched}
\begin{tabular}{llll}
\hline
 $\beta$ & Geometry & $N_{\rm config}$ & $m_q$ \\
\hline
 5.70 & $16^3\times32$&259 & .01 to .08 \\
 5.80 & $16^3\times32$&200 & .01 to .04 \\
 5.90 & $16^3\times32$&200 & .01 to .04 \\
 6.00 & $16^3\times32$&221 & .01 to .04 \\
 6.05 & $16^3\times32$&306 & .01 to .05 \\
 6.10 & $24^3\times32$& 60 & .01 to .04 \\
 6.20 & $24^3\times48$&121 & .005 to .035 \\
\hline
\end{tabular}
\end{table}

For creating kaons (at rest) we use a wall of U(1) noise on timeslice $t=0$, 
i.e. complex random numbers $\xi_{\vec{x}}$ at each space point such that
$\langle \xi_{\vec{x}}\xi^\dagger_{\vec{y}}\rangle = \delta_{\vec{x},\vec{y}}$.
This is statistically equivalent to computing a collection of delta-function
sources.  In particular, our wall creates only pseudoscalars.
\begin{figure}[htb]
\begin{center}
\leavevmode
\epsfxsize 7.2cm \epsfbox{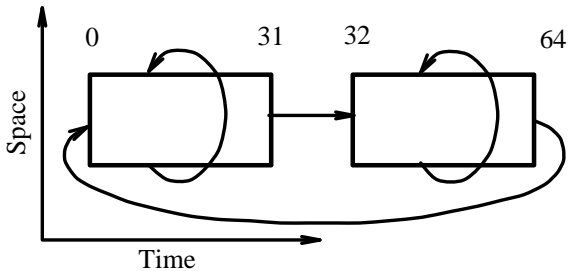}
\end{center}
\caption{We use periodic boundary conditions in space and time,
and the lattice is duplicated in time direction.}
\label{fig:period}
\end{figure}
We use a lattice duplicated in the time direction, with periodic
boundary conditions in space and time (see Fig.~\ref{fig:period}).
Computing propagators on the doubled lattice, we obtain forward-
and backward-going propagators which we use for computing $B_K$.
That is, if $G_\pi(t)$ is the $\pi$ propagator on the doubled
lattice, then our operator correlation functions are schematically
of the form $G_\pi(t)G_\pi(t+N_t)$, where $N_t=32$ or 48.

We employ three kinds of operators: Landau gauge, gauge invariant,
and tadpole improved.  Landau gauge operators are defined by
fixing the gauge and omitting explicit links in non-local operators.
For gauge-invariant  operators we supply the links, averaging over
all shortest paths.  Tadpole-improved operators are gauge-invariant
operators, but with all links rescaled by $u_0^{-\Delta}$, where
$u_0=P^{1/4}$, $P$ is the average plaquette, and $\Delta$ is the
number of links needed to connect fermion fields. 
We opted for tadpole-improved operators on all configurations,
using the others on a subset of configurations for checks.

The matching between continuum and lattice operators is of the form
$${\cal O}^{cont}_i = (\delta_{ij}+\frac{g^2}{16\pi^2}
(\gamma_{ij}\log{(\frac{\pi}{\mu a})} + C_{ij})) {\cal O}^{lat}_j, $$
where $\gamma_{ij}$ is the one-loop anomalous dimension matrix,
and $C_{ij}$ are finite coefficients, which can be sizable.
We take these from the calculations of \mbox{Refs.~\cite{IS,PS}.} 
For the continuum scheme, we choose NDR, quoting results either
at scale $\mu=\pi/a$ or at $\mu= 2\GeV$.
We use the $\overline{MS}$ coupling constant $g_{\overline{MS}}$,
defined as
$1/g^2_{\overline{MS}}(\pi /a) = P/g_{\rm bare}^2+0.02461-0.00704\, N_f.$
\begin{figure}[htb]
\begin{center}
\leavevmode
\epsfxsize 7.2cm \epsfbox{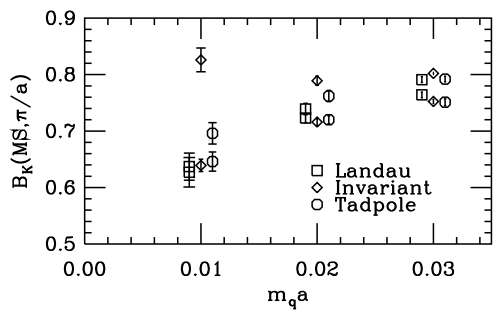}
\caption{$B_K$ with (lower points) and without (upper points)
one-loop perturbative matching.
The points are artificially displaced horizontally for clarity. }
\label{fig:pert2}
\end{center}
\end{figure}
To check how well the perturbation theory works, we computed all
three operators on a subset of the $N_f=2$ ensemble, finding that
after one-loop corrections are put in, the matrix elements
agree within our statistical error.
For the bulk of the calculation we used tadpole-improved operators
exclusively.

\begin{figure}[htb]
\begin{center}
\leavevmode
\epsfxsize 7.2cm \epsfbox{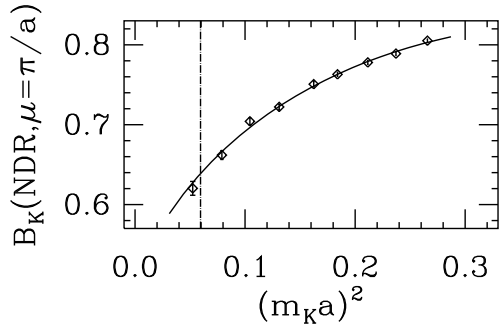}
\caption{Data and fit for $B_K$ vs. $m_K^2$ on the quenched ensemble.
The vertical line marks the physical kaon mass.}
\label{fig:BKQ}
\end{center}
\end{figure}

\section{Results for $N_f$ Dependence}

\begin{figure}[htb]
\begin{center}
\leavevmode
\epsfxsize 7.2cm \epsfbox{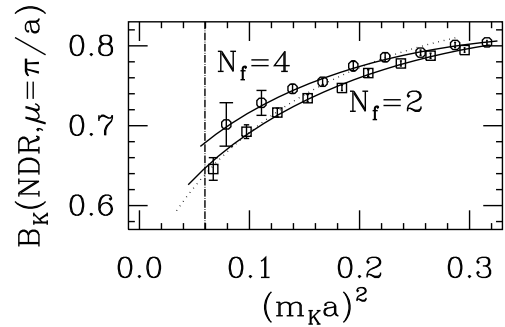}
\caption{Data and fit for $B_K$ vs. $m_K^2$ on two dynamical 
ensembles. The dashed line shows the fit for the quenched
ensemble.}
\label{fig:BKD}
\end{center}
\end{figure}

Figs.~\ref{fig:BKQ}~and~\ref{fig:BKD} show the results for $B_K$
on three ensembles of configurations. Values at 9 quark mass
points are fitted to the form expected from chiral perturbation
theory, $B_K=a+bm_K^2+cm_K^2\ln{m_K^2}$.
The \mbox{$N_f=4$} and \mbox{$N_f=2$} curves are similar in shape,
while the quenched curve crosses between the other two.
While this is perfectly allowed, we should also inject
a small note of caution---our ensembles have the same $\rho$-masses,
but these masses are presumably affected to some degree by the
finite volume.  If this effect is sizable and depends significantly
on $N_f$, our curves could shift a little.

Taking the results at face value, we note that the $N_f=2$
and $N_f=0$ results lie nearly on top of each other at the
kaon mass, consistent with our earlier results \cite{Kilcup93}.
Also, most of the $N_f=2$ data lie below $N_f=0$, consistent
with the observation by other groups that quenching seems to
increase $B_K$ slightly (see, e.g. ref. \cite{Soni}).
However, the $N_f=4$ data turn this picture upside down.
\begin{figure}[htb]
\begin{center}
\leavevmode
\epsfxsize 7.2cm \epsfbox{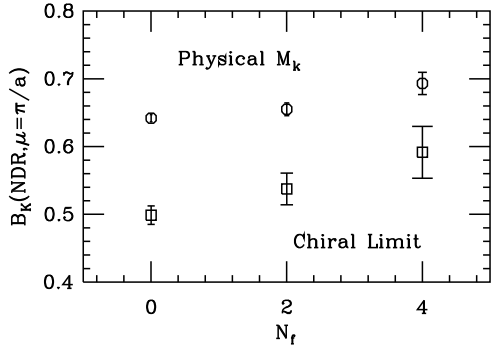}
\end{center}
\caption{Final results for $B_K$ at physical kaon mass and in
the chiral limit, vs. $N_f$.}
\label{fig:BKvsNf}
\end{figure}
Fig.~\ref{fig:BKvsNf} shows our final values for $B_K$,
obtained at the physical kaon mass and by extrapolation to
the chiral limit.  We see that the interpolated $N_f=3$ result
is a few percent higher than quenched.

\section{Continuum Extrapolation}
\begin{figure}[htb]
\begin{center}
\leavevmode
\epsfxsize 7.2cm \epsfbox{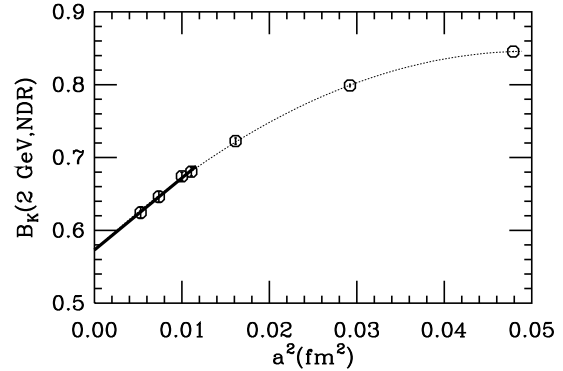}
\end{center}
\caption{Linear (heavy line) and quadratic fits to $B_K(a)$.}
\label{fig:BKvsa}
\end{figure}

Performing the same analysis on the quenched ensembles, we
obtain the result shown in figure \ref{fig:BKvsa}, where
we plot $B_K(NDR,\mu=2\GeV)$ versus the scale as determined from $m_\rho$.
The data are well fit by the quadratic form
$B_K(a)= B_K(a=0) + (a\Lambda_2)^2 + (a\Lambda_4)^4$, where
the scale of the power correction parameters turns out to
be typical of QCD: $\Lambda_2\approx660\MeV$, $\Lambda_4\approx650\MeV$.
Alternatively, we note that we can avoid making reference to the
possibly problematic $m_\rho$ by using the scaling form
$$
a(\beta) = a_0 {\big({16 \pi^2\over 11 g^2}\big)}^{51\over121}
\exp({-8 \pi^2\over 11 g^2})
$$
where we take $g$ here to be the $\overline{MS}$ coupling.
This amounts to shuffling around the $a^4$ corrections,
and in practice tends to straighten the data out.  That is
to say, much of the curvature in figure \ref{fig:BKvsa}
might be ascribed to scaling violations in $m_\rho$ itself.
To quote a final value we make the conservative choice
of a linear fit to the four points with $\beta\ge6.0$,
and obtain
$$B_K|_{a=0,N_f=0} = .573\pm.015.$$

\section{Conclusions}
From the dynamical comparison, we find that $B_K(N_f=3)$
is ($5\pm2$)\% larger than $B_K(N_f=0)$.
Combining with the $a=0$ extrapolation we
we quote our current central value $B_K$ in
the real world:
$$B_K(NDR,\mu=2\GeV,N_f=3,a=0) = .60\pm.02$$
Remaining uncertainties include possible finite-size effects
in the dynamical ensemble, higher order perturbative corrections
in the matching, and higher order chiral ($m_s-m_d$) effects.
A study of hadronic weak matrix elements relevant for $\epsilon^\prime
/\epsilon$ using the same techniques and ensembles is currently underway.

\bigskip
\noindent{\bf Acknowledgements.} We thank the Columbia collaboration
for access to the dynamical configurations.  Cray T3D time was
supplied by the Ohio Supercomputer Center and the Los Alamos
Advanced Computing Laboratory.


\begin{thebibliography}{99}
\bibitem{GKPS1} G. Kilcup, S. Sharpe, R. Gupta and A. Patel, 
{\it Phys. Rev. Lett.} {\bf 64} (1990) 25.
\bibitem{GKPS2} S. Sharpe, R. Gupta and G, Kilcup, {\it Nucl. Phys. 
 B (Proc. Suppl.)} 26 (1992) 197.
\bibitem{GKPS3} S. Sharpe, {\it Nucl. Phys. B (Proc. Suppl.) 34}
(1994) 403.
\bibitem{Ishizuka93} N. Ishizuka {\it et al.}, {\it Phys. Rev. Lett.} 
71 (1993) 24.
\bibitem{Lee} M. Klomfass and W. Lee, Lattice 1995, p. 469. 
\bibitem{Aoki95} JLQCD Collaboration, Lattice 1995, p. 465.
\bibitem{Aoki96} JLQCD Collaboration, these proceedings.
\bibitem{Kilcup93} G. Kilcup, {\it Phys. Rev. Lett.} 
71 (1993) 1677.
\bibitem{Chen} D. Chen, these proceedings.
\bibitem{IS} N Ishizuka, Y. Shizawa, {\it Phys. Rev.} 
D49 (1994) 3519.
\bibitem{PS} S. Sharpe and A. Patel, {\it Nucl. Phys.}
B417 (1994) 307.
\bibitem{Soni} A. Soni, Lattice 95, p. 43.

\end{thebibliography}
\end{document}